\documentclass[9pt]{IEEEtran}
\usepackage{amsmath,amssymb,mathrsfs}
\usepackage{graphics}
\usepackage{graphicx}
\usepackage{subfigure}
\graphicspath{{figures/png/}}
\usepackage{amsfonts}
\usepackage{latexsym}
\usepackage{url}
\usepackage{ifthen}  
\usepackage{bbm}
\usepackage{subfigure}
\usepackage{comment}
\usepackage{color}

\newtheorem{theorem}{Theorem}

\newcommand{\ket}[1]{| #1 \>}
\renewcommand{\>}{\rangle}
\newcommand{\<}{\langle} 
\newcommand{\OUT}{|\mathrm{OUT}\>}
\newcommand{\IN}{|\mathrm{IN}\>}
\newcommand{\diracPsi}{|\Psi\>}
\newcommand{\diracPsit}{|\Psi(t)\>}

\begin{document}
\title{Structured singular value analysis for spintronics network information transfer control}
\author{Edmond A. Jonckheere,~\IEEEmembership{Life~Fellow,~IEEE},  
Sophie G. Schirmer,~\IEEEmembership{Member,~IEEE}, 
and Frank C. Langbein,~\IEEEmembership{Member,~IEEE}
\thanks{EAJ is with the Department of Electrical Engineering, 
University of Southern California, Los Angeles, CA 90089, {\tt jonckhee@usc.edu.}} 
\thanks{SGS is with the College os Science, Swansea Univesity, Swansea, Wales, UK,
{\tt s.schirmer@swansea.ac.uk}.}
\thanks{FCL is with the School of Computer Science \& Informatics,
Cardiff University, Cardiff, Wales, UK, {\tt F.C.Langbein@cs.cardiff.ac.uk.} 
EAJ was supported by ARO MURI W911NF-11-1-0268.
SGS and FCL were supported by Ser Cymru National Research Network for Advanced Engineering and
Materials (grant NRN 082).}
\thanks{Manuscript received October 2016; revised March 2017.}
}
\maketitle

\begin{abstract}
Control laws for selective transfer of information encoded in 
excitations of a quantum network, based on shaping the energy landscape 
using time-invariant, spatially-varying bias fields, can be successfully 
designed using numerical optimization. 
Such control laws, already departing from classicality by replacing 
closed-loop asymptotic stability with alternative notions of localization,   
have the intriguing property 
that for all practical purposes they achieve the upper bound on the fidelity, 
yet the (logarithmic) sensitivity of the fidelity 
to such structured perturbation as spin coupling errors and bias field leakages is nearly vanishing. 
Here, these differential sensitivity results are extended to large 
structured variations using $\mu$-design tools 
to reveal a crossover region in the space of controllers where objectives usually thought to be conflicting are actually concordant.  
\end{abstract}

\begin{IEEEkeywords}
quantum mechanics, spin polarized transport, spintronics, robust control, sensitivity analysis, uncertain systems
\end{IEEEkeywords}

\section{Introduction}

Spintronics networks~\cite{QuantumSpintronicsReview,SpintronicsReview} are characterized  by their unique property that information is encoded in spin excitation (understood as ``spin up"),  
so that information transfer is mediated through spin coupling and occurs without charge movement. 
This offers more efficient information transfer than in devices that move charges, 
as heat dissipation is not a limiting factor. 
Excitation transport could occur through intrinsic spin dynamics, but with poor fidelity.  
Here, fidelity is understood as the overlap, in the sense of the 
absolute value of the inner product, 
between some desired wave function encoding a specific spin ``up" and the actual terminal wave function. 
The point that is demonstrated here is that, 
despite {\it large} uncertainties, fidelity can be brought very near to its upper bound of $1$ by means of controls  
taking the form of localized bias magnetic fields. 
This approach contrasts with another approach based on engineering the couplings.  
Whatever the approach, the technological challenges to be overcome 
are still significant. 
In particular, the bias magnetic fields can only be focused with limited resolution and  
the couplings can only be engineered with limited precision. For this reason, it is essential to assess 
how sensitive, {\it how robust,} such control schemes are against coupling and field focusing errors. 

As observed in~\cite{Edmond_IEEE_AC}, 
the static bias magnetic field controllers that nearly achieve the fidelity upper bound have nearly vanishing (logarithmic) sensitivity to model uncertainties.  
This is a puzzling observation given that traditionally control 
predicts that one cannot simultaneously achieve small logarithmic sensitivity to uncertain parameters 
and small error, here defined as the departure from the fidelity upper bound. This puzzle, explained in~\cite{statistical_control}, 
already points to quantum control deviating from classical control, 
in that the latter requires a physical measurement feedback process 
whereas the former can be accomplished by field mediation. 

Here, we bring one more component to this broader study: 
robustness against {\it large} rather than differential uncertainties 
using $\mu$-analysis of modern robust multivariable theory~\cite{Zhou}.  
By the same token, the $\mu$-analysis also reveals robustness against initial state preparation error. 
Both traditional sensitivity and $\mu$-function are consistent in that they both define a crossover 
region where controllers start losing their fidelity with accrued sensitivity, 
while in the same region the $\mu$-increases. 
These observations reinforce the earlier conclusions of~\cite{Edmond_IEEE_AC}.



\subsection{Outline}

The basic concepts associated with the control of spintronics networks for {\it time-domain} 
high fidelity information transfer are reviewed in Section~\ref{s:basic_concepts}. 
As a preview to the robustness issue,  in Section~\ref{s:sensitivity}, we explore
the (differential) sensitivity of the achieved fidelity to coupling uncertainties and bias field leakage errors. 
``Sensitivity" means that the perturbations are infinitesimal. 
From this point on, the perturbations become larger and we examine ``robustness." 
In Section~\ref{s:diagonal_perturbation}, the perturbation is formulated as a diagonally structured feedback.  
In Section~\ref{s:time_versus_frequency}, the ground work is prepared 
for the reformulation of the time-domain figure of merit to its frequency-domain counterpart.  
In Section~\ref{s:mu_analysis}, the $\mu$-analysis is set up, and in Section~\ref{s:11_spin_example}, 
the $\mu$-analysis is performed on an 11-spin ring, providing controllers maintaining good fidelity under large perturbations. 

\section{Basic concepts}
\label{s:basic_concepts}

\subsection{Information transfer control}

We consider homogeneous chains or rings 
of $N$ spins with XX or Heisenberg couplings~\cite{quantum_rome,first_with_Sophie,chains_QINP,rings_QINP,statistical_control,time_optimal}. 
Each spin can be up $\ket{\uparrow}$ or down $\ket{\downarrow}$. 
Within the Hilbert space $C^{2^N}$ of this network, we define the single excitation subspace 
as the span of $\ket{\downarrow}^{\otimes (k-1)}\otimes \ket{\uparrow}\otimes \ket{\downarrow}^{\otimes ( N-k)}$,  
where $k$ runs from $1$ to $N$. Intuitively, this is the subspace where exactly one spin is ``up." 
This subspace is invariant under the motion and in this subspace the Hamiltonian takes the matrix representation 
\begin{equation}
\label{e:Hamiltonian}
H=\left(\begin{array}{cccccc}
0           & 1   & 0   & \ldots & 0& h_{1,N} \\
1           & 0   & 1   &            & 0& 0\\
0           & 1   & 0   &            &0 & 0\\
 \vdots            &      &  \ddots   &   \ddots         & \ddots& \vdots\\
 0           &   0   &  0   &            & 0& 1\\
h_{N,1} & 0    &   0   &  \ldots      & 1 & 0
\end{array}\right),
\end{equation}
where $h_{1,N}=h_{N,1}=0$ for XX-chains and $h_{1,N}=h_{N,1}=1$ for XX-rings. 
For Heisenberg XXX-couplings, 
an identity matrix should be added to~\eqref{e:Hamiltonian}.  
The Hilbert space of the system now is $\mathbb{C}^N$ with natural basis $\{e_k\}_{k=1}^N$. 
In this single excitation subspace, $\ket{\Psi}=e_k$ 
denotes the state where the excitation (the only ``spin up") is on spin \#k.  
To achieve the objective of {\it transporting} the excitation 
from an initial wave function state $|\mathrm{IN}\>=\ket{\Psi(0)}$, localized at a spin, 
to a terminal state $|\mathrm{OUT}\>=\ket{\Psi(t_f)}$, localized at another spin, 
spatially distributed but time-invariant 
bias fields $\{D_i:i=1,...,N\}$ are applied to the respective spins. 
Defining $D=\mbox{diag}\{D_i:i=1,...,N\}$  
results in the controlled Hamiltonian $H+D$.  
With this Hamiltonian, 
and under the assumption that the system is isolated from its environment, 
the evolution is described by 
Schr\"odinger's equation $|\dot{\Psi}(t)\>=-\imath (H+D)|\Psi(t)\>$. 
Schr\"odinger's equation can be broken down, a bit  artificially, into a classical state space equation
\begin{equation}
 |\dot{\Psi}(t)\>=-\imath H|\Psi(t)\>+u(t),  
\label{e:Schrodinger}
\end{equation}
driven by the control 
\begin{equation}
\label{e:control}
u(t)=-\imath D |\Psi(t)\>. 
\end{equation}
Eqs.~\eqref{e:Schrodinger}-\eqref{e:control} formulate the transport  
problem in the set-up of control theory, with the drawback 
that the ``feedback"~\eqref{e:control} is only ``virtual." 

In~\cite{time_optimal,Edmond_IEEE_AC}, 
for each $(\IN,\OUT)$ pair, a set of bias controllers $\{D(m)\}_{m=1}^{1000}$, 
where $D(m)=\mathrm{diag}(D_{1}(m),D_{2}(m),...,D_{N}(m))$, was derived  
and the controllers were 
ordered by decreasing order of their squared fidelities, or probabilities of achieving successful transfer,  
\[  p_{t_f(m)}(\IN,\OUT)=  \left|\left\< \mathrm{OUT}|e^{-\imath (H+D(m))t_f(m)}\IN\right\>\right|^2,\] 
where $t_f(m)$ is the time at which the controller $D(m)$ achieves its maximum fidelity. 
Such fidelity or probability of successful transfer will be referred to as ``instantaneous." 
The design was a purely numerical approach to the problem of achieving optimal fidelity 
\begin{equation}
\label{e:objective}
\max_D | \< \mathrm{OUT} | \Psi(t_f) \> |  \leq 1
\end{equation}
 in a minimum amount of time $t_f$. 
The latter is motivated by the need to act faster than the decoherence. 
Because the landscape in which the optimization is performed is extremely challenging~\cite[Fig. 2]{time_optimal}, 
only {\it some} runs were successful at getting very close to the upper bound of $1$, while other runs were not as successful, 
with the reward that this gave us controllers achieving various level of fidelity, 
opening the road to explore potential conflicts with other objectives such as sensitivity and robustness.  

\subsection{Quantum-classical control design discrepancies}


\subsubsection{Measurements or no measurements?}
\label{s:no_measurements}

Eqs.~\eqref{e:Schrodinger}-\eqref{e:control} certainly allow us to formulate the quantum mechanical 
problem as a control problem, 
that is, a state-space equation~\eqref{e:Schrodinger} driven by a control $u$, 
itself depending  linearly on the state $\ket{\Psi}$ as seen by~\eqref{e:control}. 
The problem is that breaking the Schr\"odinger equation into two parts introduces a control $u$ 
that is artificial, that does not have physical existence, but that nevertheless exists mathematically. 
Even though there is no {\it physical} closed-loop backward measurement signal flow, 
the ``virtual" feedback structure, even somewhat ``hidden," has been shown to  
endow the system with good differential sensitivity properties 
relative to spin coupling uncertainties~\cite{statistical_control,Edmond_IEEE_AC}. 
This is a property certainly consistent with measurement feedback. 

One of the purposes of this paper is to clarify whether those differential sensitivity properties translate to  robustness under larger variations of the spin couplings. 


Attempts at classifying the many quantum control laws abound, 
but here we will particularly retain the classification of~\cite{tarn_SICON}, 
which emphasizes, as the present paper does,  
time-invariant spatially-distributed control. 
Control is defined in~\cite{tarn_SICON} as manipulating matter-field or field-field interactions. 
In this classification, our approach is rather a {\it field-field} interaction, 
or in other words our controller is {\it field-mediated.}  
The localized magnetic fields that are applied to the spins  
change the total energy of the system through the spin-field interaction;  
those magnetic fields are registered in the Hamiltonian as the additional diagonal terms of $D$, 
which alter the dynamics so to achieve a specific transport.  


\subsubsection{Lack of closed-loop stability}

Eq.~\eqref{e:control} is a bit misleading, as the control $D$ is \emph{selective,} in the sense that it depends on both $\IN$ and $\OUT$. 
This is contrary to classical tracking where the initial state is the quiescent state and the controller is designed to go to any terminal state. 
The selectivity implies that we need to repudiate the classical closed-loop stability. Indeed, if a controller is designed so as to achieve 
$\IN \to \OUT $, no other initial state say $\ket{\mathrm{IN}'}$ would reach $\OUT$ even asymptotically; 
indeed, because of the unitary evolution $\|U(t_f)(\IN-\ket{\mathrm{IN'}})\|=\| \IN-\ket{\mathrm{IN'}}  \|\ne 0$. 
Besides, it is obvious that the closed-loop matrix $-\imath (H+D)$  is purely oscillatory. 

In physics language, even though $\ket{\Psi(t_f)}$ might get very close to $\OUT$ at some specific $t_f$,  
over the larger time interval the best 
one can expect is to have $\ket{\Psi(t)}$ oscillate in a localized manner around $\OUT$. The oscillation is ``localized" in the sense that 
the support of the wave function is a compact neighborhood of $\OUT$. 
This is akin to the concept of Anderson localization~\cite{Anderson-58,short_to_Anderson,50_years}, 
except that Anderson localization is formally  meant to keep an excitation localized around a spin, 
whereas here it is an unexpected property of the transfer controller.

\subsection{Structured uncertainties}

The objective is to examine the robustness of the $D$-scheme under perturbation of the 2-body interaction strengths, that is, a perturbation of the Hamiltonian matrix $H$ that takes the form 
\begin{small}
\begin{align*}
&H+\Delta_H\\
&=\left(\begin{array}{ccccc}
0           & 1+\delta_{12}   & \ldots & 0 & H_{1,N}+\delta_{1,N} \\
1+\delta_{12}           & 0   &            & 0& 0\\
             &      &                 & & \\
 0           &   0   &              & 0& 1+\delta_{N-1,N}\\
H_{N,1}+\delta_{1,N} & 0    &         & 1+\delta_{N-1,N} & 0
\end{array}\right)\\
&= H+\sum_{k=1}^{N-1} \delta_{k,k+1}S_{k,k+1}+\delta_{1,N}S_{1,N},
\end{align*}
\end{small}
where $S_{k,k+1}$ is the {\it structure} associated with the perturbation of the $(k,(k+1))$ coupling, 
with the convention that  $\delta_{1,N}=0$ if $h_{1,N}=0$ (chain), and $\delta_{k,k+1}$ 
is the magnitude of the perturbation. 
Clearly, this is a {\it structured} perturbation, 
and the classical way of assessing robustness against such a structured perturbation is via the $\mu$-analysis~\cite{complex-analytic,real_versus_complex,Zhou}. 

Other uncertainties to be taken into consideration are the inaccuracies in the localization of the magnetic fields implementing the biases. 
In this case, the Hamiltonian is perturbed as 
\[ H+D+\Delta_D=H+D+\sum_{k=1}^N \delta_{kk} S_{kk} D_k, \]
where 
\[ S_{kk}=\mathrm{diag}\left(0,0,...,0,\frac{1}{2},-1,\frac{1}{2},0,...,0,0\right), \]
assuming that the bias that should nominally be applied to spin $k$ spill over symmetrically to the nearest-neighbor spins. 


\section{Small perturbation sensitivity}
\label{s:sensitivity}

\begin{figure}
\begin{center}
\subfigure{\scalebox{0.4}{\includegraphics{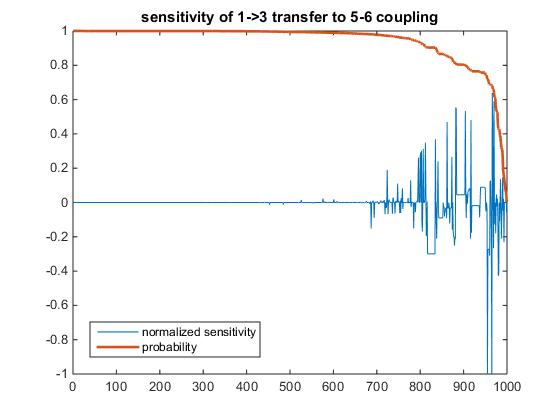}}} 
\subfigure{\scalebox{0.4}{\includegraphics{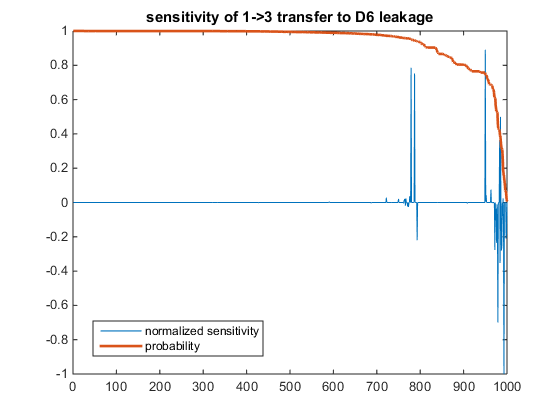}}} 
\end{center}
\caption{Plot of 
instantaneous probability (squared fidelity) and sensitivity versus index $m$ of the controllers; 
top: sensitivity relative to coupling uncertainty; bottom: sensitivity relative to leakage of bias field to near neighbor spins.}
\label{f:fidelity_vs_sensitivity}
\end{figure}

\begin{figure}
\begin{center}
\subfigure{\scalebox{0.3}{\includegraphics{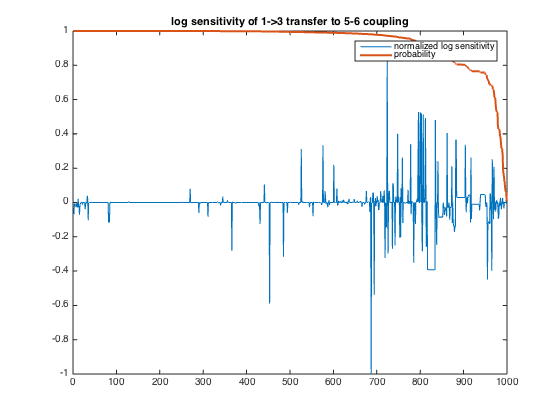}}} 
\subfigure{\scalebox{0.3}{\includegraphics{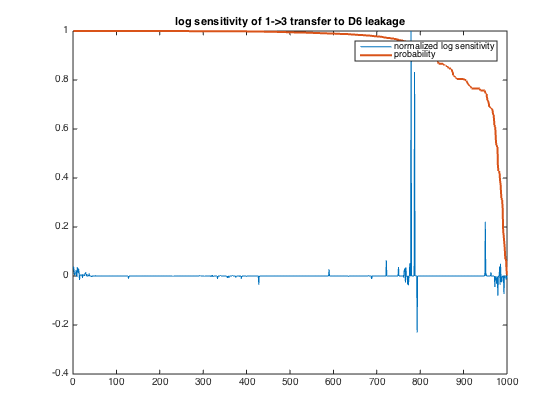}}} 
\end{center}
\caption{Plot of instantaneous probability (fidelity squared) 
and logarithmic sensitivity versus index $m$ of the controllers; 
top: log sensitivity relative to coupling uncertainty; bottom: log sensitivity relative to leakage of bias field to near neighbor spins.}
\label{f:fidelity_vs_logsensitivity}
\end{figure}

The fundamental objective in~\cite{time_optimal,Edmond_IEEE_AC} was 
the maximization  of the squared fidelity 
or probability, 
$\left|\left\< \mathrm{OUT}|e^{-\imath (H+D(m))t_f(m)}\IN\right\>\right|^2$,  
over $D$-structured controllers, 
{\it for the nominal model of the spin network.}  
This raises the question of how sensitive the squared fidelity is to network modeling errors 
and departure of the controller from bias fields highly localized around their respective spins: 
\begin{equation} 
\label{e:differential_sensitivity}
\begin{split}
\frac{\partial}{\partial \delta_{k,k+1}} 
\left|\left\< \mathrm{OUT}|e^{-\imath (H+D(m)+\delta_{k,k+1}S_{k,k+1})t_f(m)}\IN\right\>\right|^2, \\
\frac{\partial}{\partial \delta_{kk}} 
\left|\left\< \mathrm{OUT}|e^{-\imath (H+D(m)+\delta_{k,k}S_{k,k}D_k(n))t_f(m)}\IN\right\>\right|^2.
\end{split}
\end{equation}
Classically, one would expect the fidelity and the inverse sensitivity (a measure of ``robustness") 
 to go in opposite direction. 
As shown by Fig.~\ref{f:fidelity_vs_sensitivity}, this is not the case, 
as the best fidelity controllers have the least sensitivity. 
From both plots of Fig.~\ref{f:fidelity_vs_sensitivity}, the sensitivity deteriorates (increases) 
as soon as the squared fidelity begins to dip. 
The same pattern holds for the logarithmic sensitivity, where the expressions in 
Eqs.~\eqref{e:differential_sensitivity} are divided by 
the error, that is, $1-|\<\mathrm{OUT}|\Psi(t_f(m))\>|^2$, as shown in Fig.~\ref{f:fidelity_vs_logsensitivity}.  
Both plots indicate a ``crossover" region where fidelity and sensitivity begin to change markedly. 
One of the purposes of the paper is to show that the $\mu$-analysis is consistent with this finding.

\section{Large diagonally-structured perturbation}
\label{s:diagonal_perturbation}

\subsection{Coupling uncertainty}

Consider the closed-loop perturbed system
\begin{align}
\label{e:perturbed}
|\dot{\Psi}(t)\>=-\imath (H+D)|\Psi(t)\>
&-\imath\sum_{k=1}^{N-1} \delta_{k,k+1}S_{k,k+1}\diracPsit \nonumber\\
&-\imath \delta_{1,N}S_{1,N}\diracPsit .
\end{align}
The objective of maximization of $|\<\mathrm{OUT}\diracPsit|$ is easily seen to be equivalent to minimization of 
$|\<\mathrm{OUT}^\perp\diracPsit|$, where $|\mathrm{OUT}^\perp\>$ denotes a basis of the orthogonal complement of 
$|\mathrm{OUT}\>$. The output signal that usually assesses the performance of a control system can hence 
be defined rather classically as 
\[ z(t)=\<\mathrm{OUT}^\perp\diracPsit = C \Psi(t), \]
%
where the rows of the matrix $C$ form a basis of $\OUT^\perp$. 
Contrary to the classical control paradigm where a disturbance signal drives the system, 
here the system responds to an initial state $\IN$. 
With the time-invariant bias matrix $D$, 
the output $\diracPsit$ and the control $u(t)$ 
can be ``virtually" connected via $-\imath D$, as shown in Fig.~\ref{f:feedback}. 


\subsubsection{One single coupling uncertainty}
\label{s:zetav}

Computation of robustness margin against structured uncertainties relies on extracting  
the uncertain parameters from the perturbed system and displaying them  
as a diagonally structured ``fictitious" feedback from an artificially 
defined output $\zeta$ to an artificially defined input $v$ both defined on the unperturbed system, 
as shown in Fig.~\ref{f:feedback}. 
To unravel this structure, we use the matrix inversion lemma:
\begin{eqnarray*}
\lefteqn{\left( sI+\imath H +\imath \delta_{k,k+1}S_{k,k+1}\right)^{-1}=}\\
&&(sI+\imath H)^{-1}\\
&-&(sI+\imath H)^{-1} \delta_{k,k+1}I\left( I+\imath S_{k,k+1}(sI+\imath H)^{-1}\delta_{k,k+1}I\right)^{-1}\\
&&\times\imath S_{k,k+1}(sI+\imath H)^{-1}. 
\end{eqnarray*}
Observe that the uncertainty $\delta_{k,k+1}$ must be multiplied by the identity $I_{N \times N}$ to safeguard compatibility among the sizes of the various matrices, something that unfortunately 
creates a significant curse of dimensionality. Now, write the open-loop unperturbed plant as 
\[ \left(\begin{array}{c}
\zeta\\
\diracPsi  
\end{array}\right)=
\left(\begin{array}{cc}
P_{11} & P_{13} \\
P_{31} & P_{33}
\end{array}\right) 
\left(\begin{array}{c}
v \\
u
\end{array}\right). \]
A fictitious feedback $\Delta$ from $\zeta$ to $v$ would give the transfer matrix from $u$ to $v$ as
\[
(sI+\imath H +\imath \delta_{k,k+1}S_{k,k+1})^{-1}=
P_{33}+P_{31}\Delta (I-P_{11} \Delta)^{-1} P_{13}. \]
Comparing the two expressions for the open-loop system perturbed by the fictitious feedback yields
\[ \begin{small}\left(\begin{array}{cc}
P_{11} & P_{13}\\
P_{31} & P_{33}
\end{array}\right)=
\left(\begin{array}{ccc}
-\imath S_{k,k+1}(sI+\imath H)^{-1}   & \imath S_{k,k+1}(sI+\imath H)^{-1} \\
-(sI+\imath H)^{-1} & (sI+\imath H)^{-1}
\end{array}\right) \end{small}\]
together with
\[ \Delta =\delta_{k,k+1}I_{N\times N}. \]

With $P_{11}$, $P_{13}$, $P_{31}$ and $P_{33}$ taken care of, 
it remains to define the second block row of $P$ (output variable $z$) and the second block column of $P$ 
(input variable $\Psi(0)$). The idea is to observe that the second and last block columns of $P$ are the same, since $u$ and $|\mathrm{IN}\>$ have exactly the same effect on the dynamics. 
From there on, making use of the matrix $C$, it is easily seen that 
$(P_{21},P_{22},P_{23})=C(P_{31},P_{32},P_{33})$. Then  
we derive the remaining second block row and second block columns  of $P$ as
\begin{eqnarray*}
P_{32}&=&P_{33}, \\
P_{21}&=&CP_{31}, \\
P_{22}&=&CP_{32}, \\
P_{23}&=&P_{22}, \\
P_{12}&=&P_{13}. 
\end{eqnarray*}
Setting $\Phi:=(sI+\imath H)^{-1}$ to simplify the notation, the block $3 \times 3$ 
plant equation of Fig.~\ref{f:feedback} becomes
\begin{equation}
\label{e:1Pmatrix}
\left(\begin{array}{c}
\zeta\\
z\\
\Psi
\end{array}\right)
\left(\begin{array}{rrr}
-\imath S_{k,k+1} \Phi & \imath S_{k,k+1} \Phi & \imath S_{k,k+1} \Phi\\
-C\Phi & C \Phi & C\Phi\\
-\Phi & \Phi & \Phi
\end{array}\right)
\left(\begin{array}{c}
v\\
\IN\\
u
\end{array}\right).
\end{equation}

\subsubsection{Many uncertain couplings}

To avoid clutter, we consider only two coupling uncertainties, $\delta_{k,k+1}$ and $\delta_{\ell,\ell+1}$. 
The general pattern of $N$ (or $N-1$) uncertain couplings for a ring (or a chain) will clearly emerge from this simple case. It is claimed that the subset of $P$-equations relevant to the uncertainty feedback model is
\[ \left(\begin{array}{c}
\zeta_k \\
\zeta_\ell\\
\Psi
\end{array}\right)=
\left(\begin{array}{rrr}
-\imath S_{k,k+1}\Phi & -\imath S_{k,k+1}\Phi & \imath S_{k,k+1}\Phi\\
-\imath S_{\ell,\ell+1}\Phi & -\imath S_{\ell,\ell+1}\Phi & \imath S_{\ell,\ell+1}\Phi\\
-\Phi & -\Phi & \Phi
\end{array}\right)
\left(\begin{array}{c}
v_k\\
v_\ell \\
f
\end{array}\right), \]
where $f:=\IN + u$ is the forcing term. Using the matrix inversion lemma, it is easily seen that closing the loop 
$v_k=\delta_{k,k+1}I_{N\times N}\zeta_k$ yields
\[ \left(\begin{array}{c}
\zeta_\ell\\
\Psi
\end{array}\right)=
\left(\begin{array}{rr}
-\imath S_{\ell,\ell+1}\Phi_k & \imath S_{\ell,\ell+1}\Phi_k\\
-\Phi_k &  \Phi_k
\end{array}\right)
\left(\begin{array}{c}
v_\ell \\
f
\end{array}\right), \]
where $\Phi_k:=(sI+\imath H+\imath \delta_{k,k+1} S_{k,k+1})^{-1}$. Appealing one more time to the matrix inversion lemma reveals that closing the loop $v_\ell=\delta_{\ell,\ell+1}I_{N\times N}\zeta_\ell$ yields
\begin{align*} 
\Psi &= \left( \Phi_k^{-1} + \delta_{\ell,\ell+1} \imath S_{\ell,\ell+1} \right)^{-1}f \\
     &=(sI+\imath \delta_{k,k+1} S_{k,k+1} +\imath \delta_{\ell,k+1} S_{k,k+1})^{-1} f,
\end{align*}
as claimed. 

In this case of two uncertain couplings, the block $3 \times 3$ 
plant equation becomes
\[\left(\begin{array}{cc|c|c}
-\imath S_{k,k+1}\Phi & -\imath S_{k,k+1} \Phi & \imath S_{k,k+1}\Phi & \imath S_{k,k+1} \Phi\\
-\imath S_{\ell,\ell+1}\Phi & -\imath S_{\ell,\ell+1} \Phi & \imath S_{\ell,\ell+1}\Phi 
                                                & \imath S_{\ell,\ell+1} \Phi\\\hline
-C\Phi & -C\Phi & C\Phi & C\Phi \\\hline
-\Phi & -\Phi & \Phi & \Phi  
\end{array}\right). \]

\subsection{Leakage of bias field to near-neighbor spins}

In case of uncertainty on the focusing power of the bias field, 
restricting ourselves to the nominal bias field $D_{kk}$ 
spilling over spins $k-1$ and $k+1$, the perturbed dynamics becomes
\begin{equation}
\label{e:perturbed_leak}
|\dot{\Psi}(t)\>=-\imath (H+D)|\Psi(t)\>
-\imath\sum_{k=1}^{N} \delta_{k,k}S_{k,k}D_{kk}\diracPsit. 
\end{equation}

\subsection{General architecture}
\label{s:general_architecture}

\begin{figure}
\centerline{
\scalebox{0.4}{\includegraphics{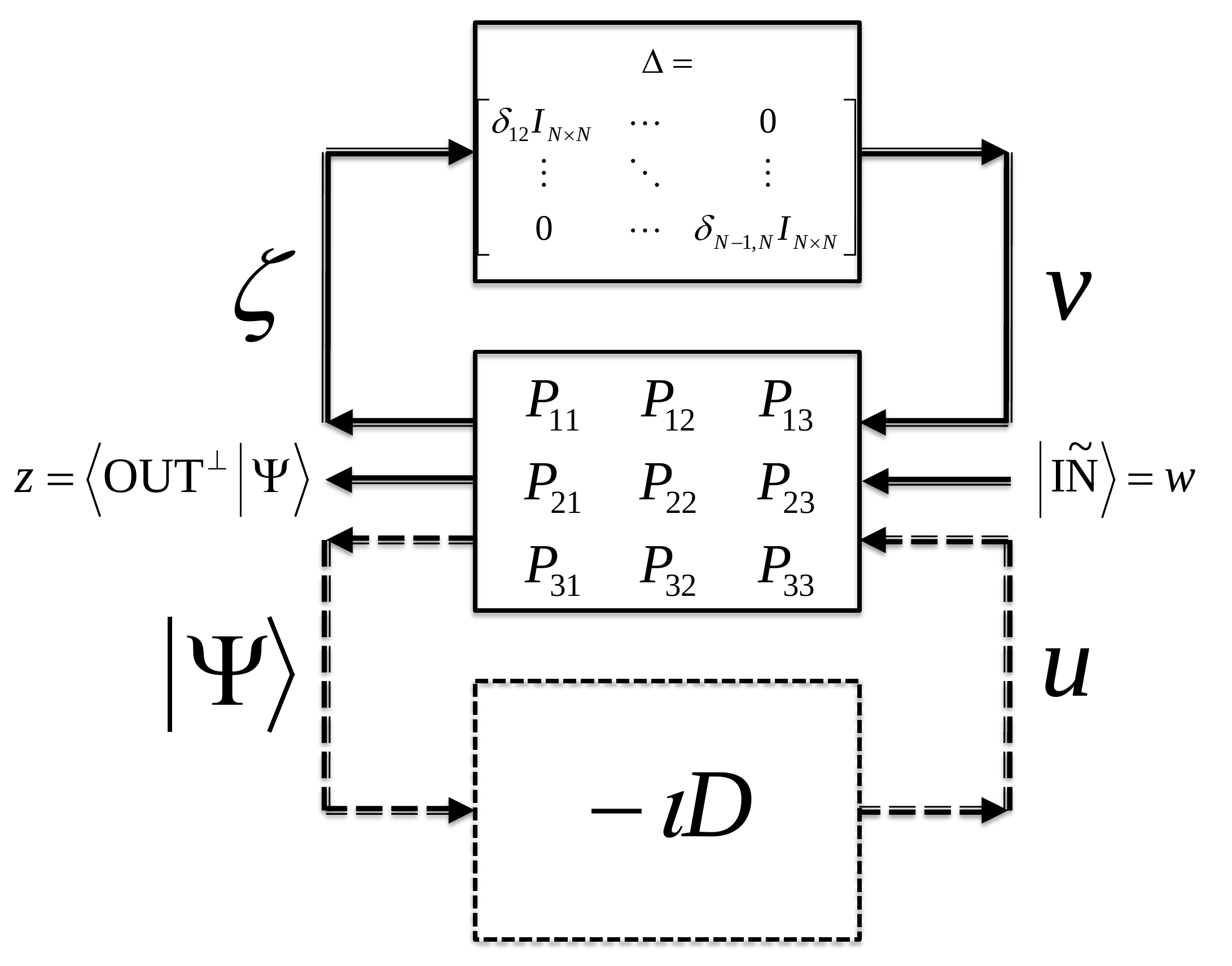}}}
\caption{Modern robust multivariable control inspired diagram showing the perturbation as a diagonally structured
feedback. The dotted path from $\ket{\Psi}$ to $u$ means that the feedback is virtual, 
not measurement mediated.}
\label{f:feedback}
\end{figure}

The overall control-inspired architecture of the quantum system is shown in Fig.~\ref{f:feedback}. 
The $3 \times 3$ system $P$ is open-loop, unperturbed. 
The top feedback introduces the uncertain parameters, as explained in Sec.~\ref{s:zetav}. 
The bottom (dotted) flow is a representation of the control~\eqref{e:Schrodinger}-\eqref{e:control}.  
This flowchart might give 
the wrong impression that there is a need to measure $\diracPsit$ 
with the potential danger of back-action of the measurement. 
Quite to the contrary, as already argued in Sec.~\ref{s:no_measurements}, 
there is no measurement feedback, 
only a {\it virtual} feedback. 
The prepared state $\IN$ is the initial condition, 
viewed as a disturbance input $w(t)$ to the system, 
modeling a constant but uncertain preparation error on the initial state.  
The overall system is set up in such as way as to respond to $w(t)$, as shown in Fig.~\ref{f:feedback}. 
The output is the error $\<\mathrm{OUT}^\perp|\Psi(t)\>$. Since
\[ |\<\mathrm{OUT}^\perp|\Psi(t)\>|^2 + |\<\mathrm{OUT}|\Psi(t)\>|^2 =1, \]
to secure a squared fidelity $|\<\mathrm{OUT}|\Psi(t)\>|^2\geq 1-\epsilon$, it suffices to take 
$|\<\mathrm{OUT}^\perp|\Psi(t)\>|^2\leq \epsilon$.

\subsection{Time-domain versus frequency-domain design}
\label{s:time_versus_frequency}

The fidelity with the bias control $D(m)$ remains oscillating around the target state and the maximum 
fidelity is recorded at $t_f(m)$. This is of course a time-domain approach, 
with difficulties to be translated to the traditional frequency-response methods of modern robust multivariable control. 
Here we take a simplified approach to the problem by trading the maximum fidelity 
over time for a time-averaged fidelity. Besides, the instantaneous fidelty can never been observed, 
as the read-out device can only time-average over a finite window. 

\begin{figure}
\begin{center}
\scalebox{0.6}{\includegraphics{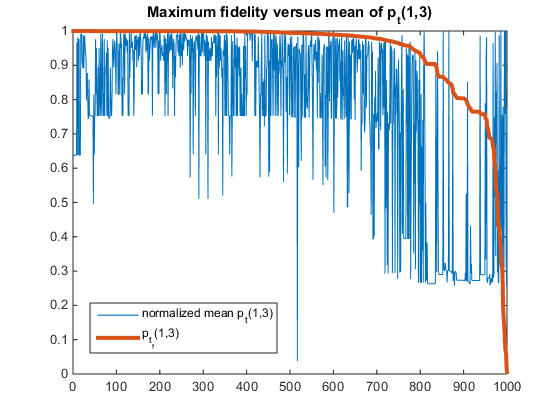}}
\end{center}
\caption{Behavior of maximum squared fidelity at $t_f(m)$  versus time-averaged fidelity}
\label{f:time_vs_frequency}
\end{figure}

The question of whether this time-average can be related to the time-optimal fidelity 
is here justified only by a simulation study. 
In Fig.~\ref{f:time_vs_frequency}, the controllers are ordered by decreasing value of the fidelity at $t_f(m)$ 
and compared with an average fidelity 
obtained by a Lyapunov method (which correlates very well with the averaging over $2 t_f(m)$). 
A qualitative concordance between the behavior of the mean fidelity and 
that of the time-optimal fidelity can be seen, confirmed by a Kendall $\tau$ of 0.3621.

\section{$\mu$-analysis}
\label{s:mu_analysis}

In~\cite{time_optimal}, given an $(\IN,\OUT)$ pair, 
a collection $\{D(m)\}_{m=1}^{1000}$ of diagonal matrices 
satisfying the transport requirement in a varying amount of time with varying degree of success was derived.  
The objective here is to assess the fidelity versus the $\mu$-robustness. 
Classically, one would expect a trade-off between fidelity and sensitivity.  
However, Fig.~\ref{f:fidelity_vs_sensitivity} indicates that this well known fundamental limitation in its differential form 
does not survives the passage to the quantum world. 
The problem is that no matter how encouraging the near vanishing sensitivity mediated by the best fidelity controllers is, 
it does not answer the question of large deviation, as the sensitivity in~\cite{Edmond_IEEE_AC} was computed for $\delta=0$. 
We attempt to address this question classically  
using the robust performance formalism~\cite[Chap. 10]{Zhou}, 
but we immediately have to deviate from the classical formalism on two major points:
\begin{enumerate}
\item There is no stability requirement.  
Indeed, the wave function of the physically motivated Anderson localization~\cite{Anderson-58} is purely oscillatory.
\item The disturbance input $w(t)$ of Fig.~\ref{f:feedback} is not in $L^2$, but is   
a constant in time but uncertain initial preparation $\tilde{\IN }$ of $\IN$. 
In other words, using a concept developed in~\cite{Belevitch}, 
the input is in exponential regime $e^{0t}$. 
Hence that part of the output at that exponential regime is 
\[ \widehat{z}(0)=C\widehat{\Psi}(0)=C\left.(sI+\imath(H+D))^{-1}\right|_{s=0}\tilde{\IN }.\]
With the convention that $w=\ket{\tilde{\mathrm{IN}}}$, this leads us to define
\[  T_{zw}(s)=\left.C(sI+\imath(H+D(\IN,\OUT))^{-1}\right|_{s \approx 0}  \]
as the closed-loop performance indicator. Indeed, setting $\tilde{\mathrm{IN}}=\mathrm{IN}+\boldsymbol{\delta}$, 
we obtain for near perfect state transfer controllers $T_{zw}\ket{\tilde{\mathrm{IN}}}\approx T_{zw}\boldsymbol{\delta}$, 
so that $T_{zw}w$ provides the response to the preparation error amplified by the uncertainty lumped in $T_{zw}$. 
\end{enumerate}

\begin{figure}[t]
\centerline{
\scalebox{0.4}{\includegraphics{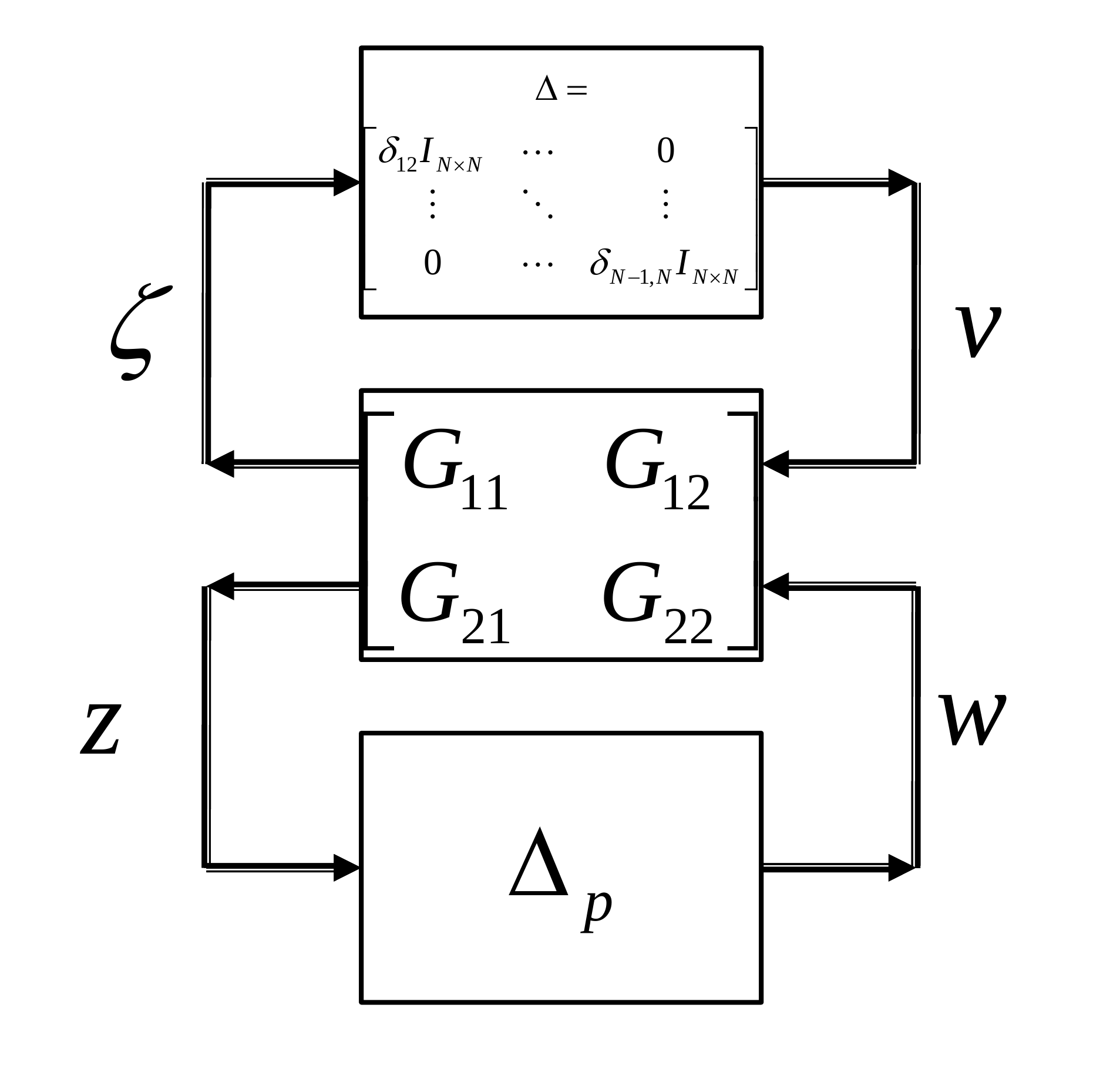}}}
\caption{Classically inspired robust performance design showing a fictitious perturbation $\Delta_p$ feedback from performance output to disturbance input}
\label{f:feedback12}
\end{figure}

We proceed from Fig.~\ref{f:feedback} and absorb the controller in the plant to come up with
\begin{equation}
\label{e:G}
\left(\begin{array}{c}
\zeta\\
z
\end{array}\right)=
\left(\begin{array}{cc}
G_{11} & G_{12}\\
G_{21} & G_{22}
\end{array}\right)
\left(\begin{array}{c}
v\\
w
\end{array}\right),
\end{equation}
where
\begin{equation}
\label{e:Gmatrix}
\begin{split}
&G_{11}=P_{11}-P_{13}\imath D (I+P_{33}\imath D)^{-1} P_{31},\\
&G_{12}=P_{12}-P_{13}\imath D(I+P_{33} \imath D)^{-1} P_{32},\\
&G_{21}=P_{21}-P_{23}\imath D(I+P_{33} \imath D)^{-1}P_{31},\\
&G_{22}=P_{22}-P_{23}\imath D(I+P_{33} \imath D)^{-1}P_{32},
\end{split}
\end{equation}
as shown in Fig.~\ref{f:feedback12}. 
From Eq.~\eqref{e:G}, it is easily seen that the performance can be assessed via 
\[ T_{zw}=G_{22}+G_{12}\Delta\left( I-G_{11}\Delta \right)^{-1}G_{12}. \]

The $\mu$-design specification (see, e.g.,~\cite[Th. 10.8]{Zhou}) 
is formulated as 
\begin{equation}
\label{e:specification}
 \|T_{zw}\|\leq \beta \mbox{   for all   } \|\Delta\| < 1/\beta. \
\end{equation}
As $\beta$ decreases, observe that the response to the initial preparation error decreases 
while this decrease occurs for larger uncertainties. Hence $\beta$ is a robustness measure. 
As we are soon to show, $\beta$ can be evaluated via the $\mu$-function. 

The specification~\eqref{e:specification} can be rewritten as $\det(I-T_{zw} \Delta_p) \ne 0$ for all fictitious perturbations  
$\|\Delta_p\|<1/\beta$. The fictitious perturbation can be reinterpreted as a feedback from $z$ to $w$ 
and the specification $\|T_{zw}\|\leq \beta$ for all $\|\Delta_p\| < 1/\beta$ means that  
$\det(I-T_{zw}\Delta_p)\ne 0$. The difficulty is that $T_{zw}$ depends on another $\Delta$. This difficulty 
can be overcome by observing that 
\begin{equation}
\label{e:big_blocks}
\det(I-G_{11}\Delta)\det\left( I-T_{zw}\Delta_p \right)= \det(I- G \mathbf{\Delta}), 
\end{equation}
where $\mathbf{\Delta}$ is the augmented structured perturbation
\[ \mathbf{\Delta}=\left(\begin{array}{cc}
\Delta & 0\\
0 & \Delta_p
\end{array}\right), \quad \|\mathbf{\Delta}\| < 1/\beta. \]
%
$\mathbf{\Delta}$ is structured in two different ways: first of all it has block diagonal structure and secondly 
$\Delta$ is diagonal. Ideally, the structure imposed upon $\Delta_p$ 
is that its column space should reproduce the space of initial preparations of $\ket{\Psi(0)}$. 

In view of~\eqref{e:big_blocks}, the condition for robust performance now becomes 
$\det(I-G\mathbf{\Delta})\ne 0$ for all structured $\|\mathbf{\Delta}\| < 1/\beta$. 
It is well known (see, e.g.,~\cite[Th. 10.8]{Zhou}) that the latter is equivalent to 
\[ \mu_{\mathcal{D}}(G)\leq \beta, \]
where $\mu_{\mathcal{D}}$ denotes the $\mu$-function, or structured singular value, relative to the  set of matrices $\mathcal{D}$  
structured as $\mathbf{\Delta}$. Specifically, 
\[ \mu_{\mathcal{D}}(G)=\frac{1}{\min\{\|\mathbf{\Delta}\|: \mathbf{\Delta} \in \mathcal{D}, \det(I-G\mathbf{\Delta})=0\}}. \]

Observe that the upper bound $\beta$ can be interpreted as $\emph{some}$ sensitivity of the performance relative to the structured perturbation. Indeed, if $\beta$ is small, $\|T_{zw}\|$ remains small for large $\Delta$'s and hence the sensitivity is small. An exact relationship, if any, between $\beta$ and the sensitivity in the sense of Sec.~\ref{s:time_versus_frequency}, Eq. \eqref{e:differential_sensitivity}, is hard to come by; however, the next section will provide an illustration that such a relationship is plausible. 

\section{11-spin ring simulation example}
\label{s:11_spin_example}

Here we give an example illustrating that the unconventional behavior 
of the fidelity versus spin coupling sensitivity  
depicted in  Figures~\ref{f:fidelity_vs_sensitivity}-\ref{f:fidelity_vs_logsensitivity} 
survives the double passage (i) from instantaneous to time-averaged performance 
and (ii) from differential sensitivity to robustness against larger variations---and initial preparation errors. 

We take an 11-ring with nominally uniform coupling strengths between near neighbor spins, 
but subject to a 5-6  uncertain coupling, under $|1\>\to |3\>$ transfer. 
For this transfer, we have the controllers $\{D(m)\}_{m=1}^{1000}$ initially ordered by decreasing value of $p_{t_f(m)}(1,3)$ at our disposal. 
The time-average of the probability is computed using a Lyapunov method; 
this defines the permutation $I(\cdot)$ of $\{1,...,1000\}$ such that $I(m)$ is the rank of the controller $D(m)$ in 
the new classification by decreasing order of time-averaged probability; 
hence  the controllers are reordered as $\{D(I(m))\}_{m=1}^{1000}$  
 by decreasing value of the time-averaged transfer probability they achieve.  
Naturally, as already observed from Fig.~\ref{f:time_vs_frequency}, one cannot expect complete consistency 
between the performance at the best time and the time-averaged performance. Nevertheless, despite this discrepancy, we show some unconventional behavior of the robust design $\mu_\mathcal{D}(G)$ versus the time-averaged performance. 
Traditionally, one would expect the robustness to deteriorate ($\mu \uparrow$) 
as the performance improves 
($|\<\mathrm{OUT}|\Psi\>| \uparrow$), 
but we show that around some controllers the reverse behavior happens. 

To be somewhat more precise as to how the simulation was performed, we took 
an $11 \times 11$ Hamiltonian of the form~\eqref{e:Hamiltonian} with $h_{1,11}=h_{11,1}=1$. 
The $P$-matrix of Fig.~\ref{f:feedback} was taken as in Eq.~\eqref{e:1Pmatrix}. In this  $P$-matrix, 
the output matrix $C$ defining $z=C\Psi $
was taken as 
$C=\left(\begin{array}{cccccc}
e_1 & e_2 & 0_{11,1} & e_4 & \ldots & e_{11}
\end{array}\right)^T$, where $\{e_k\}_{k=1}^{11}$ is the natural basis of $\mathbb{C}^{11}$ over $\mathbb{C}$ and $0_{11,1}$ is an $11$-dimensional column vector with $0$s everywhere; the matrix $S_{k,k+1}$ was taken as the $11 \times 11$ matrix with zeros everywhere except for $1$'s in positions 
$(5,6)$ and $(6,5)$; the matrix $\Phi$ was taken as $(sI+\imath H)^{-1}_{s=0}=(\imath H)^{-1}$.   
The $G$-matrix of Fig.~\ref{f:feedback12} was computed as in Eq.~\eqref{e:Gmatrix}. The $\mathcal{D}$ block structure was defined 
$\delta I_{11 \times 11} \oplus \mathbb{C}^{11\times 11}$. The $\mu_{\mathcal{D}}(G$) was computed using the {\tt mussv} function of Matlab.  

Note that the $\OUT$ selectivity is picked up by the output $z$, but the analysis falls short of the $\IN$  
selectivity. The latter would require a $\Delta_p$ matrix structured as a single row in position corresponding to the $\IN$ spin, but this appears beyond the capability of the {\tt mussv} at this stage. 

\begin{figure}
\begin{center}
\subfigure{\scalebox{0.6}{\includegraphics{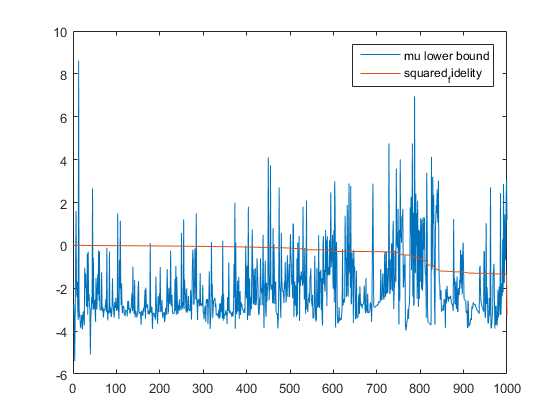}}}
\subfigure{\scalebox{0.6}{\includegraphics{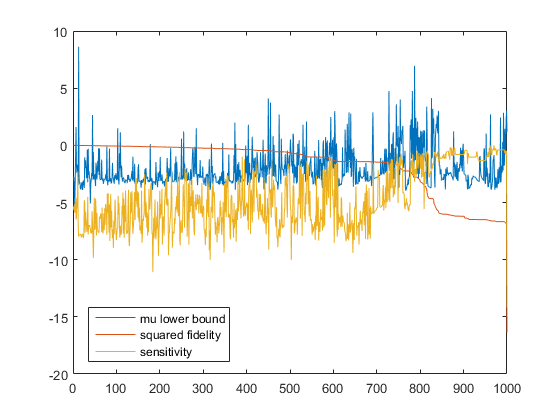}}}
\end{center}
\caption{Robust design $\mu$, average fidelity, and differential sensitivity achieved by various controllers ordered by decreasing value of their average fidelity 
(bottom plot was rescaled for better visualization.)}
\label{f:mu_vs_fidelity}
\end{figure}

The results are summarized in Fig.~\ref{f:mu_vs_fidelity}. On the top panel, especially around controllers $I(m)\in[700,850]$, one notices the unconventional behavior of decrease of the performance (fidelity $\downarrow$) concomitant with decrease of the robustness ($\mu \uparrow$). 
More precisely, it appears that the increase in $\mu$ follows the rate of decrease of the fidelity. 
A theoretical explanation of this latter phenomenon remains to be formulated, though. 
The bottom panel is essentially the same as the top, except for some rescaling, 
with the addition of the differential sensitivity as defined by Eq.~\eqref{e:differential_sensitivity}. 
The unconventional behavior of the differential sensitivity versus the fidelity is quite obvious, 
but more importantly observe that the differential sensitivity has its ``crossover" in the same interval, 
$I(m)\in [700,850]$, as already singled out on the top panel.

All of the above qualitative observations can be confirmed by quantitative Kendall $\tau$ analysis. 
The visually obvious consistent behavior of the differential sensitivity and the $\mu$ is confirmed by a Kendall $\tau$ of 0.6672. 
The overall Kendall $\tau$ of the $\mu$ lower bound and the squared fidelity is -0.1601, indicating a slight negative correlation, as claimed. 
In the $I(m)\in[700,850]$ area, this result can be improved by considering 
the {\it incremental} squared fidelity and $\mu$, which negatively correlate with a Kendall $\tau$  of -0.1970. 

\section{Conclusion}

We have shown that the reassuring good sensitivity properties  
of high fidelity excitation transport controllers for spintronic networks 
demonstrated in~\cite{statistical_control,Edmond_IEEE_AC} 
can be extended to larger variation, 
so that such controllers can  be objectively referred to as ``robust." 
Both classical differential sensitivity and $\mu$ analyses reveal a 
nonclassical crossover region in the space of controllers where fidelity,  
classical sensitivity, and robustness as quantified by the $\mu$-function all deteriorate---a  
rather surprising observation that contradicts the classical limitations on achievable performance.  
This quantum-classical discrepancy can be explained by the field-mediation 
of the quantum control while classical control is measurement-mediated. 

It is still technologically challenging to address the individual spins of the network with well 
focused magnetic fields.  
This is the reason why sensitivity against field focusing errors was considered in~\cite{Edmond_IEEE_AC} 
and shown to vanish for perfect state transfer controllers.    
This technological difficulty could be overcome on a theoretical basis 
by addressing ``islands" of spins instead of single spins 
using multiple excitation~\cite{k_excitation_subspace} 
and possibly next nearest neighbor couplings~\cite{Heisenberg_with_next_near_neighbor}.  
This is left for further research. 
A more technically viable solution is provided by ultracold atom quantum simulators~\cite{cold_atoms}, 
where laser pulses can address single atoms 
and the energy landscape can be controlled, as it is here. 
More specifically, Heisenberg chain simulators based on 
optical lattices~\cite{2016_Heisenberg_simulator}, 
atom cavities~\cite{1D_Heisenberg_spin_chain_simulation}, 
trapped ions~\cite{trapped_ions_Heisenberg_simulator_journal}  
offer plenty of techniques for energy landscape shaping and single atom pseudo-spin addressing that reproduce 
the theoretical model adopted here  
(with a preference for cavity models where single atom addressing seems easier~\cite{1D_Heisenberg_spin_chain_simulation}). 
Furthermore, some copper compounds~\cite{Heisenberg_Copper} 
nearly reproduce the theoretical behavior of the 1-dimensional chains considered here.  

We have chosen controllers that transport the excitation in a minimum amount of time~\cite{time_optimal} to mitigate decoherence. 
Conceptually, the approach proposed here can easily be extended to networks subject to decoherence by replacing Schr\"odinger's equation with von Neumann-Lindblad's equation, 
but at the expense of a serious curse of dimensionality, 
as the original $N$-dimensional problem now becomes a $N^2$-dimensional problem. 
This is  left for further research.

\bibliographystyle{plain}
\bibliography{../../../../../../bookstore/coarse,../../../../../../bookstore/chaos,../../../../../../bookstore/oxford,../../../../../../bookstore/networking,../../../../../../bookstore/edmond,../../../../../../bookstore/physics,../../../../../../bookstore/power_grid}

\begin{thebibliography}{10}

\bibitem{Anderson-58}
P.~W. Anderson.
\newblock Absence of diffusion in certain random lattices.
\newblock {\em Physical Review}, 109(5):1492--1505, March 1958.

\bibitem{Belevitch}
V.~Belevitch.
\newblock {\em Classical Network Theory}.
\newblock Holden-Day, San Francisco, 1968.

\bibitem{cold_atoms}
Immanuel Bloch, Jean Dalibard, and Sylvain Nascimb\`ene.
\newblock Quantum simulations with ultracold quantum gases.
\newblock {\em Nature Physics}, 8:267–276, 2012.

\bibitem{1D_Heisenberg_spin_chain_simulation}
Z.-X Chen, Z.-W Zhou, X.~Zhou, X.-F Zhou, and G.-C. Guo.
\newblock Quantum simulation of {H}eisenberg spin chains with next nearest
  neighbor interactions in coupled cavities.
\newblock {\em Phys. Rev. A}, 81:022303, 2010.
\newblock arXiv:1212.5328v1 [quant=ph] 21 Dec 2012.

\bibitem{tarn_SICON}
W.~B. Dong, R.-B.Wu, W.~Zhang, C.~W. Li, and T.~J. Tarn.
\newblock Spatial control model and analysis of quantum fields in
  one-dimensional waveguides.
\newblock {\em SIAM Journal on Control and Optimization}, 54(3):1352--1377,
  2016.

\bibitem{QuantumSpintronicsReview}
D.~D.~Awschalom \emph{et~al.}
\newblock Quantum spintronics: Engineering and manipulating atom-like spins in
  semiconductors.
\newblock {\em Science}, 339(6124):1174--1179, March 2013.

\bibitem{SpintronicsReview}
S.~A.~Wolf \emph{et~al.}
\newblock Spintronics: A spin-based electronics vision for the future.
\newblock {\em Science}, 294(5546):1488--1495, November 2001.

\bibitem{trapped_ions_Heisenberg_simulator_journal}
T.~Grass and M.~Lewenstein.
\newblock Trapped-ion quantum simulation of tunable-range {H}eisenberg chains.
\newblock {\em EPJ {Q}uantum {T}echnology}, 1:8, 2014.
\newblock doi:10.1140/epjqt8.

\bibitem{short_to_Anderson}
Dirk Hundertmark.
\newblock A short introduction to {A}nderson localization.
\newblock In {\em Proceedings of the {LMS} Meeting on Analysis and Stochastics
  of Growth Processes and Interface Models}, pages 194--218. Oxford University
  Press, Oxford, United Kingdom, 2008.

\bibitem{2016_Heisenberg_simulator}
C.-L. Hung, A.~Gonzales-Tudela, J.~Ignacio Cirac, and H.~J. Kimble.
\newblock Quantum spin dynamics with paiwise-tunable, long-range interactions.
\newblock {\em PNAS}, pages E4946--E4955, August 2016.

\bibitem{quantum_rome}
E.~Jonckheere, F.~C. Langbein, and S.~G. Schirmer.
\newblock Curvature of quantum rings.
\newblock In {\em Proceedings of the 5th International Symposium on
  Communications, Control and Signal Processing (ISCCSP 2012)}, Rome, Italy,
  May 2-4 2012.
\newblock {DOI}: 10.1109/{ISCCSP}.2012.6217863.

\bibitem{first_with_Sophie}
E.~Jonckheere, S.~Schirmer, and F.~Langbein.
\newblock Geometry and curvature of spin networks.
\newblock In {\em IEEE Multi-Conference on Systems and Control}, pages
  786--791, Denver, CO, September 2011.
\newblock DOI: 10.1109/CCA.2011.6044395. Available at arXiv:1102.3208v1
  [quant-ph].

\bibitem{chains_QINP}
E.~Jonckheere, S.~Schirmer, and F.~Langbein.
\newblock Quantum networks: {T}he anti-core of spin chains.
\newblock {\em Quantum Information Processing}, 13:1607--1637, 2014.
\newblock Published on line May 24, 2014. (DOI: 10.1007/s11128-014-0755-5).
  Available at {\tt http://eudoxus2.usc.edu}.

\bibitem{rings_QINP}
E.~Jonckheere, S.~Schirmer, and F.~Langbein.
\newblock Information transfer fidelity in spin networks and ring-based quantum
  routers.
\newblock {\em Quantum Information Processing ({QINP})}, 14(10), 2015.
\newblock DOI: 10.1007/s11128-015-1136-4; available at {\tt
  http://eudoxus2.usc.edu} and arXiv:submit/1359959 [quant-ph] 24 Sep 2015.

\bibitem{statistical_control}
E.~Jonckheere, S.~Schirmer, and F.~Langbein.
\newblock Jonckheere-{T}erpstra test for nonclassical error versus
  log-sensitivity relationship of quantum spin network controllers.
\newblock {\em International Journal of Robust and Nonlinear Control}, 2016.
\newblock Submitted, available at arXiv:1612.02784 [math.OC].

\bibitem{complex-analytic}
E.~A. Jonckheere and Nainn-Ping Ke.
\newblock Complex-analytic theory of the $\mu$-function.
\newblock {\em Journal of Mathematical Analysis and its Applications},
  237:201--239, 1999.

\bibitem{real_versus_complex}
E.~A. Jonckheere and Nainn-Ping Ke.
\newblock Real versus complex robustness margin continuity as a smooth versus
  holomorphic singularity problem.
\newblock {\em Journal of Mathematical Analysis and its Applications},
  237:541--572, 1999.

\bibitem{Heisenberg_with_next_near_neighbor}
L.~C. Kweck, Y.~Takahashi, and K.~W. Choo.
\newblock Spin chain under next nearest neighbor interaction.
\newblock {\em Journal of Physics: Conference Series}, 143, 2009.
\newblock doi: 10.1088/1742-6596/143/1/012014.

\bibitem{50_years}
A.~Lagendijk, B.~van Tiggelen, and D.~Wiersma.
\newblock Fifty years of {A}nderson localization.
\newblock {\em Physics Today}, 62:24--28, August 2009.

\bibitem{time_optimal}
F.~Langbein, S.~Schirmer, and E.~Jonckheere.
\newblock Time optimal information transfer in spintronics networks.
\newblock In {\em IEEE Conference on Decision and Control}, pages 6454--6459,
  Osaka, Japan, December 2015.

\bibitem{Heisenberg_Copper}
J.~M. Maillet.
\newblock Heisenberg spin chains: from quantum groups to neutron scattering
  experiments.
\newblock {\em S\'eminaire Poincar\'e}, X:129--177, 2007.

\bibitem{k_excitation_subspace}
T.~Orsborne.
\newblock Static and dynamic of quantum {XY} and {H}eisenberg systems on
  graphs.
\newblock arXiv: quant-ph/0312126v3, 2006.

\bibitem{Edmond_IEEE_AC}
S.~Schirmer, E.~Jonckheere, and F.~Langbein.
\newblock Design of feedback control laws for spintronics networks.
\newblock {\em IEEE Transactions on Automatic Control}, 2016.
\newblock Under revisions; available at arXiv:1607.05294.

\bibitem{Zhou}
K.~Zhou and J.~C. Doyle.
\newblock {\em Essentials of robust control}.
\newblock Prentice Hall, Upper Saddle River, NJ, 1998.

\end{thebibliography}

\end{document}